# The Competition Between Deformation Twinning and Dislocation Slip in Deformed Face-Centered Cubic Metals

Ritesh Jagatramka      Matthew Daly[*]

*Department of Civil, Materials, and Environmental Engineering, University of Illinois at Chicago – 842 W. Taylor St., 2095 ERF (MC 246), Chicago, IL, 60607, United States*

**ABSTRACT**

The competition between deformation twinning and dislocation slip underpins the evolution of mesoscale plasticity in face-centered cubic materials. While competition between these mechanisms is known to be related to the critical features of the generalized planar fault energy landscape, a physical theory that tracks competition over extended plasticity has yet to emerge. Here, we report a methodology to predict the mesoscale evolution of this competition in deformed crystals. Our approach implements kinetic Monte Carlo simulations to examine fault structure evolution in face-centered cubic metals using intrinsic material parameters as inputs. These results are leveraged to derive an analytical model for the evolution of the fault fraction, fault densities, and partitioning of plastic strains among deformation mechanisms. In addition, we define a competition parameter that measures the tendencies for deformation twinning and dislocation slip. In contrast to previous 'twinnability' parameters, our derivation considers deformation history when examining mechanism competition. This contribution therefore extends the reach of deformation twinning theory beyond incipient nucleation events. These products find direct applications in work hardening and crystal plasticity models, which have previously relied on phenomenological relations to predict the mesoscale evolution of deformation twin microstructures.

[*]Corresponding author: mattdaly@uic.edu (M. Daly)

1. **INTRODUCTION**

The mesoscale plasticity of face-centered cubic (FCC) metals is underpinned by the operation of competing deformation mechanisms. Amongst these, dislocation slip and deformation twinning are widely recognized to be two important mechanisms that actively compete during plastic deformation. The comparative dominance of one mechanism is determined by a complex interplay between intrinsic material properties and extrinsic factors. Competition in the former category can be conceptualized using the generalized planar fault energy (GPFE) landscape, which has its roots in works from Vítek [1,2]. Various investigators have leveraged the GPFE landscape concept to produce parameter-based descriptors of deformation mechanism competition. For competition between deformation twinning and slip, Tadmor and co-workers provided the seminal parameters. Their earliest work defines a twinning tendency criterion for the onset of deformation twinning at a crack-tip [3], where a direct relationship between the critical features of the GPFE landscape (*i.e.,* the unstable stacking fault and twinning energies) and deformation twinning is defined. These results demonstrate the multi-parameter dependencies of deformation twinning and challenge the general belief that twinning tendency is driven solely by the intrinsic stacking fault energy. A subsequent work broadened this approach by homogenizing the crack-tip model over a distribution of crack orientations in a polycrystal [4]. Asaro and Suresh [5] considered a specific slip system geometry, under the crack-tip parameter of Tadmor and Hai, to examine the competition between deformation twinning and dislocation slip at grain boundaries in nanostructured FCC materials. Jin et al. [6] reparameterized the criterion of Asaro and Suresh to provide a single parameter relation for twinning tendency under the original analytical framework of Tadmor and co-workers.

In an independent approach, Jo et al. [7] consolidated considerations of crystal orientation and the GPFE to develop a unified parameter that predicts tendencies for deformation twinning, slip, and stacking fault emission. These descriptors of competition between deformation twinning and slip are referred to here as 'twinnability' parameters, following the nomenclature of Tadmor and Bernstein [4]. Each of these parameters is summarized in a recent review from De Cooman et al. [8].

While these twinnability parameters provide a fundamental understanding of the intrinsic competition between deformation mechanisms, there are some notable limitations. Namely, these descriptors offer insight into incipient deformation tendencies (*i.e.,* the first emission of an extended dislocation or formation of a twin embryo from stacking of adjacent planar faults) but do not track competition as deformation proceeds. Consequently, these parameters cannot be leveraged to determine the evolution of correlated phenomena such as work hardening, which requires consideration of deformation history. Nor can they be used to predict the partitioning of plastic strain amongst the mechanisms of deformation twinning and dislocation slip. These limitations become evident in twinning-induce plasticity (TWIP) steels [9–14], where the relative contributions of deformation twinning and dislocation slip are well-known to vary over the stages of work hardening [8,15,16]. Additional systems of technological relevance, where the evolution of mechanism competition is important, include nanotwinned materials [17–19] and high entropy alloys [20–24]. Analytical efforts to segment the contributions of dislocation slip and deformation twinning in work hardening and crystal plasticity models are well documented, with significant contributions presented in the works of Bouaziz and coworkers [25–29], Kim et al. [30], Steinmetz et al. [16], and Kalidindi [31,32]. However, a shortcoming in each of these approaches is the reliance on empirical relations for the accumulation of deformation twins during deformation, which can provide aphysical results. For instance, early empirical modeling efforts estimate a twin fraction as high as 0.69 in TWIP steels [25]. Later works have predicted a twin fraction in the range of ~0.10-0.20 [27,33], with 0.15 being the commonly accepted value [8]. While these later predictions better align with experimental observations, the broad applicability of current evolution models remains poor due to their reliance on phenomenology and empirical fitting.

Within the context of twinnability, an opportunity exists to propose new, physical models that not only track the competition between deformation twinning and dislocation slip but provide predictive tools to examine deformed microstructures under extended plastic deformation. Here, we present a methodology to quantify the partitioning of plastic strain between deformation twinning and dislocation slip mechanisms and measure the accumulation of fault structures in deformed FCC crystals. For this purpose, the competition between deformation twinning and dislocation slip is studied using kinetic Monte Carlo (kMC) simulations. Based on kMC simulations, a set of analytical relations are derived that leverage the critical energies of the GPFE landscape to predict the evolution of fault structures. The outcomes of this study are two-fold. The primary result provides a new method to predict the evolution of competition between deformation twinning and dislocation slip over extended plastic deformation using only intrinsic material properties as inputs. From a fundamental perspective, this contribution expands the twinnability framework originally developed by Tadmor and coworkers [3,4] by extending its scope beyond incipient events. The second outcome is a series of relations to predict the partitioning of plastic strain between deformation twinning and dislocation slip mechanisms and the storage of fault structures over extended deformation in FCC metals. We anticipate that this product will enhance existing work hardening and crystal plasticity models, by providing first-principles-based predictions of mesoscale planar fault evolution in deformed microstructures.

## 2. METHODOLOGY

### 2.1. Kinetic Monte Carlo approach

To address the question of mechanism competition, we have implemented the relevant kinetic equations for dislocation slip and deformation twinning mechanisms following the kMC algorithm outlined in Bortz et al. [34]. The kMC simulation cell can be envisioned as a discretized FCC crystal, where the kMC relations are evaluated at each node of the mesh. The nucleation and progression of defects in this cell are considered by traversing system states that are separated by kinetic barriers. These features are well-suited to the objectives of this work, which require tracking defects over extended deformation and monitoring the relative kinetics between deformation mechanisms. A similar approach has been used to examine the competition between the process of deformation twin nucleation and deformation twin thickening in our previous work [35]. The kMC method described in this section has been implemented in Python and will be made available to the community upon publication of this work through a Github repository.

The kMC simulation cell is considered as a FCC single crystal that is initially deformation free with the <110> and <111> crystallographic axes oriented along the global $x$ and $y$ direction, respectively. The simulation cell measures $Mb_{110}$ by $Nd_{111}$ where $b_{110}$ is the magnitude of the Burger's vector of the <110> dislocation and $d_{111}$ is the interplanar distance between {111} planes. The simulation cell possesses free surfaces along the $x$ axis and periodic boundaries along the $y$ axis. A schematic of the kMC simulation cell is provided in Fig. 1a. Two different kinetic processes are evaluated in kMC simulations: partial dislocation nucleation and partial dislocation glide. The operation of each of these processes is separately considered for leading and trailing partial dislocations. Competition is assessed through the sequential activation of the relevant processes required to realize deformation twinning or dislocation slip. In order to provide an intrinsic comparison, extrinsic factors such as Schmid effects are removed, and competition is examined along a single twinning/slip system. All dislocations considered in this study are <112>-type Shockley partial dislocations. Each leading and trailing partial dislocation has a Burger's vector ($b_{112}$) with a magnitude of $a_o/\sqrt{6}$, where $a_o$ is the lattice parameter. Leading and trailing partials are assumed to be conjugate, with a 60° mixed character that acts along a line parallel to the <112> crystallographic axis (global *z*-axis). Deformation proceeds through the nucleation and glide of partial dislocations under a boundary/surface-mediated mechanism, which is described in subsequent paragraphs. The incipient nucleation event of a leading partial dislocation forms an intrinsic stacking fault (ISF). Subsequent leading partial nucleation events on adjacent {111} slip planes lead to the formation of a two-layer extrinsic stacking fault (ESF) and a multi-layer twin fault (TF). Dislocation slip proceeds through the nucleation and glide of trailing partials at locations in the simulation cell where a fault structure already exists. This process causes slip in the FCC lattice and a layer-by-layer decrement in the thickness of planar faults (*i.e.*, detwinning). Upon removal of the ISF, the lattice returns to a fault-free configuration, but is in a slipped state. To enable an intrinsic study of a single twin/slip system, cross-slip mechanisms, and detwinning and slip within the interior of fault structures are not considered.

The boundary-mediated partial dislocation emission mechanism implemented in this study can be seen as an extension of the crack-tip problem considered by Tadmor and coworkers [3,4]. However, to facilitate an intrinsic comparison of deformation mechanisms over extended plasticity we have replaced the crack-tip with a surface of equivalent nucleation sites. This treatment is inspired by the boundary-mediated twin formation mechanism that is established in the experimental literature for a diverse set of systems including TWIP steels [10], nanostructured

FCC materials [36–38], nanowires [39,40], and hexagonal close-packed metals [41]. This twin formation mechanism is distinct from classical processes such as the Cohen-Weertman [42] and Fujita-Mori [43] cross-slip mechanisms and the pole-based mechanism of Venables [44] but bears some similarities to the three-layer twin nucleus mechanism of Mahajan and Chin [45]. We have previously validated our implementation of this formation mechanism against molecular dynamics simulations of deformation twin nucleation and growth in FCC nanowires [35]. One important note regarding our approach is that dislocation processes are considered as homogeneous events, where the system is agnostic of local microstructural heterogeneities (*e.g.*, crack-tips, grain boundary energies) that may bias rates. This treatment has the intended effect of providing an intrinsic comparison of deformation mechanisms that arise explicitly from their various process barriers. Our approach is similar to that of Jo et al. [7], where a homogeneous treatment was used to study the competition between incipient mechanisms. Heterogeneities may only arise in this study due to fault structures that emerge from the deformation history. Yet, the kMC approach is sufficiently general such that microstructure heterogeneities can be specified with some effort. Although this modification is not trivial, it is not necessary to achieve the objectives of this work.

The barriers to dislocation nucleation and glide processes are defined using the energies ($\gamma$) of the GPFE landscape (see Fig. 1b), following the method of Ogata et al. [46]. In this approach, the barrier that acts at the $j^{th}$ slip plane within the crystal is determined by the local fault environment and thus reflects the deformation history of the system (see Fig. 1a). We have selected four common FCC metals (Ag, Au, Cu, and Al) for kMC simulations, for which the GPFE landscape is well-known. This selection was found to encompass the extremes in the behaviors of mechanism competition. Deformation twinning initiates with the incipient nucleation barrier ($E_1^+$) for a leading <112>-type Shockley partial dislocation. The thickening of deformation twins proceeds by overcoming additional process barriers ($E_2^+, E_3^+, E_\infty^+$) that are defined as the difference between the relevant fault (*i.e.*, $\gamma_{isf}, \gamma_{esf}, \gamma_{tf}$) and the peak energies (*i.e.*, $\gamma_{usf}^1, \gamma_{usf}^2, \gamma_{utf}^3, \gamma_{utf}^\infty$) of the subsequent defect along the GPFE landscape. Conversely, the reverse parameters ($E_1^-, E_2^-, E_3^-, E_\infty^-$) describe the process barriers for the nucleation of trailing <112>-type Shockley partial dislocations, which activate dislocation slip. The peak energies $\gamma_{usf}^1$ and $\gamma_{usf}^2$ refer to the unstable fault energies that must be overcome to form an ISF and an ESF, respectively. Similarly, the peak energies of $\gamma_{utf}^3$ and $\gamma_{utf}^\infty$ define the energies for an embryotic and thickened deformation twin, respectively. In each case, the superscript refers to the number of leading dislocations required to form the relevant fault structure. Table I provides the values for the critical energies of the GPFE landscape (*i.e.*, $\gamma_{usf}^1, \gamma_{usf}^2, \gamma_{utf}^\infty, \gamma_{isf}, \gamma_{esf} \gamma_{tf}$) used in kMC simulations. These values are obtained from density functional theory calculations using the climbing-image nudged elastic band method, as reported in Jin et al. [6]. In FCC metals, the critical energies of the GPFE landscape are known to stabilize after the formation of an ESF [46], which can be considered as a twin embryo with two adjacent twin boundaries. Therefore, the process barriers to twinning ($E_3^+$) and detwinning ($E_3^-$) of the twin embryo are determined using the approximation $\gamma_{utf}^3 \approx \gamma_{usf}^2$. The energy of the three-layer twin embryo is taken as $\approx 2\gamma_{tf}$, where $\gamma_{tf}$ is the energy of an isolated coherent twin boundary. The process barrier for twinning and detwinning at thicknesses of beyond three {111} planes is defined by $E_\infty^+$ and $E_\infty^-$, respectively. Each of these approximations are common within the community, as discussed in Jin et al. [6] and De Cooman et al. [8].

The rates ($R_{i,j}$) of nucleation and glide events are evaluated at nodes along a 2-dimensional mesh that maps to the activation sites for these dislocation processes in the slip planes of the kMC cell. Following the kMC method, these rates are determined using the Arrhenius relation:

$$R_{i,j} = R_o \exp\left\{\frac{-(\hat{\sigma}_{i,j} - \sigma_{i,j})V}{k_b T}\right\} \quad (1)$$

where $R_o$ is the pre-exponential factor (taken as the Debye frequency [47]), $V$ is the activation volume (taken as $10b_{112}^3$, as per Ramachandramoorthy et al. [48]), $k_b$ is the Boltzmann constant, and $T$ is the temperature (set at 300 K). $\hat{\sigma}_{i,j}$ and $\sigma_{i,j}$ are the process barrier and elastic stresses, respectively, that operate at the $i^{th}$ activation site in the $j^{th}$ slip plane of the kMC simulation cell. The values for $\hat{\sigma}_{i,j}$ represent the stress to nucleate a partial dislocation or the stress for glide of a partial dislocation depending on the deformation history of the kMC simulation. For instance, in a pristine simulation cell $\hat{\sigma}_{i,j}$ reduces to $\hat{\sigma}_{0,j}$, which defines the stress to nucleate a leading partial dislocation. After nucleation of a leading partial in the $j^{th}$ slip plane, $\hat{\sigma}_{0,j}$ then becomes the stress to nucleate a conjugate trailing partial (for dislocation slip) and $\hat{\sigma}_{i,j}$ is the stress required for glide of the leading partial at the $i^{th}$ activation site (taken as the Peierls-Nabarro stress, $\sigma_{PN}$, see Fig. 1a). These nucleation and glide stresses are then updated as the kMC simulation proceeds to reflect the local fault environment. Following the method of Ogata et al. [46], the undulations of the GPFE landscape are taken as a Peierls potential and may be used to directly determine the process barriers of nucleation and glide. Dislocation nucleation stresses are retrieved from the athermal limit using a harmonic approximation for the shape of process barriers (see Ref. [46]). This simple model for nucleation finds excellent agreement with benchmarking validation studies [35]. More complex nucleation models consider the elastic energy of the nucleated dislocation and the stress-dependency of the critical energies of the GPFE [49–51]. When glide is operative, the process barrier stress may be calculated from the solution to the Peierls-Nabarro problem for a partial dislocation [52]. These considerations lead to a conditional definition for the process barrier stress $\hat{\sigma}_{i,j}$:

$$\hat{\sigma}_{i,j} = \begin{cases} \dfrac{\pi E_{0,j}}{b_{112}}, nucleation\ (i = 0) \\ K_\rho \dfrac{b_{112}}{\rho} exp\left\{\dfrac{-2\pi \varsigma_{\rho(i,j)}}{\rho}\right\}, glide\ (i \neq 0) \end{cases} \quad (2)$$

where $E_{i,j}$ is the process barrier (*e.g.*, $E_{i,j} = E_1^+, E_1^-$, *etc.*) for leading or trailing dislocations in the relevant fault environment and $\varsigma_{\rho(i,j)} = \dfrac{K_\rho b_{112}^2}{4\pi^2 E_{i,j}}$ is the half-width of the dislocation core. $\rho$ is a geometric parameter that represents the distance between adjacent atomic rows along the shear direction. $K_\rho$ is an elastic constant that is defined by the shear modulus ($G$) and Poisson's ratio ($\nu$). Following the approximation of Nabarro [53], partial dislocations were assigned an edge character for glide stress calculations (*i.e.*, $\rho = \dfrac{3}{2}b_{112}$, $K_\rho = \dfrac{G}{(1-\nu)}$). This modest simplification allows the glide barrier of dislocations to be defined by a single shear stress, which is required for kMC rate determination steps in Eq. (1). The elastic constants are calculated using the method of Bacon and coworkers [54,55]. This method provides effective isotropic constants from dislocation energy factors in anisotropic media. The relevant material parameters used in all kMC calculations are provided in Table I. The effective process barrier stress (*i.e.*, $\hat{\sigma}_{i,j} - \sigma_{i,j}$) is determined by considering the additive contributions of elastic stress fields from partial dislocations stored in the kMC simulation cell. Individual stress fields are calculated using the Volterra solution to the dislocation elasticity problem for each leading and trailing partial dislocation [56]. The relevant stress tensors are rotated to align with the Burger's vectors of the respective defect (*i.e.*, ± 60° partial dislocations). Boundary effects are accounted for using the image dislocation method, which enforces a vanishing condition along free surfaces (*i.e.*, the <110> surfaces of the kMC simulation). Further details on the dislocation elasticity calculations performed in this study are provided in the online supplementary material. In addition to stresses arising from internal defects, the application of external far-field loadings can reduce the effective process barriers. The effects of far-field loadings are not specifically considered here as they exert a uniform influence on rate kinetics. However, it should be noted that our formulation is sufficiently general to include their effects along with the associated Schmid factors.

Implementation of Eqs. (1) and (2) within the kMC method enables a kinetically-weighted observation of deformation phenomena where the likelihood of deformation twinning and dislocation slip is determined by the deformation history. At each simulation step, the fault fraction ($F$) and number of faults ($N_F$) are measured from a lineal section of the simulation cell. The calculation of the fault fraction and number includes contributions from single and two-layer defects such as ISFs and ESFs, and also twin defects (TFs), which is consistent with the treatment of the deformation twinning mechanism sequence in previous works [3,4,46]. For this reason, we use the terms fault and twin interchangeably throughout the results and discussion. The simulation cell is initialized in a pristine condition and simulations are terminated once the plastic strain ($\gamma_P$) reaches 0.2. Plastic strain is calculated as a shear strain relative to the <110> and <111> crystal axes (*i.e.*, $\gamma_{xy}$ relative to the global *x* and *y* axes). Therefore, the annihilation of 60º leading and trailing partial dislocations at the exit boundary contributes increments of $\frac{\sqrt{3}}{2\sqrt{2}}|dF|_{max}$ to the cumulative plastic strain, respectively, where $|dF|_{max} = \frac{1}{N}$.

### 2.2. Analytical model

An analytical model has been developed to track the competition between deformation twinning and dislocation slip over extended plasticity. This model consists of a system of coupled equations that can be solved using standard numerical techniques. This system of equations contains no empirical fitting parameters or phenomenological constants. The intended outcome of this effort is to provide a physical model for the evolution of planar fault structures and the partitioning of plastic strains, which can be leveraged to pass mesoscale deformation information to the microstructure/continuum scales.

The evolution of the fault fraction can be quantified as the summation of the fault fraction increment $\left(\frac{dF^+}{d\gamma_P}\right)$ and decrement $\left(\frac{dF^-}{d\gamma_P}\right)$ per increment of plastic strain:

$$\frac{dF}{d\gamma_P} = \frac{dF^+}{dN_\perp}\frac{dN_\perp}{d\gamma_P} - \frac{dF^-}{dN_\perp}\frac{dN_\perp}{d\gamma_P} \tag{3}$$

where $\frac{dN_\perp}{d\gamma_P} = \left(\frac{\sqrt{3}}{2\sqrt{2}}|dF|\right)^{-1}$ is the incremental change in the number of nucleated dislocations per increment of plastic strain. The terms $\frac{dF^+}{dN_\perp}$ and $\frac{dF^-}{dN_\perp}$ are related to the probability that a nucleation event (*i.e.*, an increment to the number of nucleated dislocations, $dN_\perp$) results in an increase or decrease in the fault fraction, respectively. These terms may be determined directly from the probabilities for leading ($P^+$) and trailing nucleation ($P^-$):

$$\frac{dF^+}{dN_\perp} = P^+|dF|_{max} \tag{4a}$$

$$\frac{dF^-}{dN_\perp} = P^-|dF|_{max} \tag{4b}$$

By inspection, the nucleation probabilities define the likelihood of an incremental or decremental change to the fault fraction (*i.e.*, $|dF|_{max} = \frac{1}{N}$) and by definition $P^+$ and $P^-$ fall in the range of 0 to 1. The fault fraction, therefore, evolves at partial increments of $-\frac{1}{N} \leq dF \leq \frac{1}{N}$, which reflects the weightings of leading and trailing nucleation probabilities. The probabilities for these events may be derived from the proportion of leading ($R^+$) and trailing ($R^-$) nucleation rates to the total rates ($R$) summed over all nucleation sites in the kMC model:

$$P^+ = \frac{R^+}{R} = \frac{(1 - F - 2n_F)\bar{E}_1^+ + 2\sum_k n_k \bar{E}_{k+1}^+}{(1 - F - 2n_F)\bar{E}_1^+ + 2\sum_k n_k (\bar{E}_{k+1}^+ + \bar{E}_k^-) - n_1 \bar{E}_1^-} \tag{5a}$$

$$P^- = \frac{R^-}{R} = \frac{2\sum_k n_k \bar{E}_k^- - n_1 \bar{E}_1^-}{(1 - F - 2n_F)\bar{E}_1^+ + 2\sum_k n_k (\bar{E}_{k+1}^+ + \bar{E}_k^-) - n_1 \bar{E}_1^-} \tag{5b}$$

where $\bar{E}_k^{+,-}$ is the relevant leading or trailing barrier coefficient such that $\bar{E}_k^{+,-} = exp\left\{\frac{-V\pi}{k_B T b_{112}} E_k^{+,-}\right\}$, and $n_k = \frac{N_k}{N}$ and $n_F = \frac{\sum_k N_k}{N} = \frac{N_F}{N}$ are the fault number densities. $n_k$ represents the density of single ($k = 1$, ISFs) and multi-layer faults ($k > 1$, ESFs and TFs) and $n_F$ is the total fault number density. Faults thicker than three layers are counted towards $n_\infty$ and evaluated using the barriers $E_\infty^+$ and $E_\infty^-$, which is consistent with kMC methods. The second term (third term) in the numerator (denominator) of Eq. (5b) avoids double counting of trailing nucleation from an ISF, which only has one nucleation site (see Fig. 1a). For further details on the derivation of Eq. (5), the reader is referred to our earlier work [35], which examines the competition between deformation twin nucleation and thickening using a related approach.

The partitioning of plastic strain amongst the mechanisms of deformation twinning and dislocation slip can be determined using the parameters defined for the evolution of the fault fraction. The dislocation slip strain ($\gamma_S$) is incremented through the operation of trailing dislocations, with the probability of these events defined by $P^-$. The increment to the dislocation slip strain is counted as twice the partial strain increment, to account for the prior operation of the leading partial that then contributes to the slip mechanism. The deformation twinning strain ($\gamma_F$) can then be calculated from the difference of the total plastic and dislocation slip strains, which leads to the following set of relations:

$$\frac{d\gamma_S}{dN_\perp} = 2P^- \frac{d\gamma_P}{dN_\perp} = P^- \sqrt{\frac{3}{2}} |dF|_{max} \tag{6a}$$

$$\gamma_F = \gamma_P - \gamma_S \tag{6b}$$

Evaluation of Eqs. (3)- (6) requires a series of evolution rules for the fault number densities. We consider two outcomes that can alter the number of faults – namely, the nucleation of leading and trailing partial dislocations. As in Eq. (3), the evolution in the number of faults with plastic strain is described by an additive relation:

$$\frac{dn_F}{d\gamma_P} = \frac{dn_F^+}{dN_\perp}\frac{dN_\perp}{d\gamma_P} - \frac{dn_F^-}{dN_\perp}\frac{dN_\perp}{d\gamma_P} \tag{7}$$

where $\frac{dn_F^+}{dN_\perp}$ and $\frac{dn_F^-}{dN_\perp}$ are related to the probabilities of an increase or decrease in the fault number density, respectively. To model the probability of an increase in the number of faults, we consider the comparative kinetics of these outcomes as deformation proceeds. That is, an increase to $n_F$ only occurs when a leading partial dislocation is nucleated in a defect-free area of a crystal and a decrease to $n_F$ is accompanied by the nucleation of a trailing partial dislocation at an ISF. The relevant quantities are defined as follows:

$$\frac{dn_F^+}{dN_\perp} = \frac{1}{N}\frac{(1 - F - 2n_F)\bar{E}_1^+}{(1 - F - 2n_F)\bar{E}_1^+ + 2\sum_k n_k(\bar{E}_{k+1}^+ + \bar{E}_k^-) - n_1\bar{E}_1^-} \tag{8a}$$

$$\frac{dn_F^-}{dN_\perp} = \frac{1}{N}\frac{2n_1\bar{E}_1^-}{(1 - F - 2n_F)\bar{E}_1^+ + 2\sum_k n_k(\bar{E}_{k+1}^+ + \bar{E}_k^-) - n_1\bar{E}_1^-} \tag{8b}$$

In order to solve Eqs. (8a) and (8b), a rule for the evolution of is $n_k$ required, which is described by the following relations:

$$\frac{dn_1}{dN_\perp} = \frac{1}{N}\frac{(1 - F - 2n_F)\bar{E}_1^+ + 2n_2\bar{E}_2^- - 2n_1\bar{E}_2^+ - n_1\bar{E}_1^-}{(1 - F - 2n_F)\bar{E}_1^+ + 2\sum_k n_k(\bar{E}_{k+1}^+ + \bar{E}_k^-) - n_1\bar{E}_1^-} \quad k = 1 \tag{9a}$$

$$\frac{dn_k}{dN_\perp} = \frac{1}{N} \frac{2n_{k-1}\bar{E}_k^+ + 2n_{k+1}\bar{E}_{k+1}^- - 2n_k\bar{E}_{k+1}^+ - 2n_k\bar{E}_k^-}{(1 - F - 2n_F)\bar{E}_1^+ + 2\sum_l n_l(\bar{E}_{l+1}^+ + \bar{E}_l^-) - n_1\bar{E}_1^-} \quad k = 2,3 \tag{9b}$$

$$n_\infty = n_F - n_1 - n_2 - n_3 \tag{9c}$$

which accounts for the change in the number densities of the various fault structures due to fault nucleation/thickening or detwinning/slip processes.

The ratio of the probabilities for leading and trailing partial dislocation nucleation is also of interest, given the influence that these parameters have on mechanism competition. We define here the competition parameter ($\eta$) as the ratio of leading and trailing probabilities as:

$$\eta = ln\left(\frac{P^+}{P^-}\right) = ln\left(\frac{(1 - F - 2n_F)\bar{E}_1^+ + 2\sum_k n_k \bar{E}_{k+1}^+}{2\sum_k n_k \bar{E}_k^- - n_1 \bar{E}_1^-}\right) \tag{10}$$

Examination of Eq. (10) offers interesting insights. Leading dislocation nucleation is favored when $\eta > 0$ and trailing dislocation nucleation is favored when $\eta < 0$. Therefore, this simple criterion enables facile tracking of deformation tendencies over extended plasticity. Given the relevance of leading/trailing nucleation to the mechanisms of deformation twinning and dislocation slip, this parameter may also be viewed as a measure of mechanism competition. Indeed, the relevant process barriers (*e.g.*, $E_1^+, E_2^+, E_1^-$ and $E_2^-$) in the competition parameter contain the GPFE parameters (*i.e.*, $\gamma_{usf}^1$, $\gamma_{usf}^2$, $\gamma_{isf}$) that are found in many of the incipient twinnability parameters available in the literature [3,4,6,7]. In addition to these variables, additional parameters appear (*i.e.*, $F, n_k$, and $n_F$) that account for the evolution of the mesoscale defect structures during deformation. In this regard, this competition parameter combines two distinct components - intrinsic material properties and microstructure parameters – to examine the evolution of mechanism competition from its measure of nucleation preferences for leading/trailing dislocations.

## 3. RESULTS AND DISCUSSION

The kMC model described in Section 2.1 is implemented to study the mechanism competition in a variety of crystal sizes for Ag, Au, Cu, and Al. The results for kMC systems measuring $60b_{110}$ by $200d_{111}$ for Ag, Au, and Cu form the basis of the analysis presented in the main text. kMC simulations of Al do not exhibit deformation twinning but are provided in the online supplementary material to demonstrate how the kMC model also captures pure dislocation slip behavior. Each kMC simulation condition has been replicated 500 times for statistical sampling and was observed to converge well below the replication limit (refer to online supplementary material). All error bars are reported as $\pm 1$ standard deviation. All contour plots are normalized from a binning strategy performed on the kMC simulation results. The raw kMC data is binned along the *x* and *y* axes of the plot and bin counts are presented as normalizations of the maximum measurements along the plot ordinate. The coupled analytical equations from Section 2.2 are sequentially solved using the 4$^{th}$ order Runge-Kutta numerical method. Additional results examining kMC size effects are provided in the online supplementary material.

### 3.1. Mesoscale evolution of deformed microstructures

Representative snapshots of the kMC simulation cell at several stages of plastic deformation are shown in Fig. 2 for Ag, Au, and Cu. The shaded stroke represents regions where a fault is currently present. As anticipated, several fault structures (*i.e.*, ISFs, ESFs, and TFs) are progressively nucleated and annihilated during plastic deformation. Ag exhibited the highest storage of faults, whereas faults nucleated in Cu were found to be rapidly annihilated by the emission of a trailing partial dislocation. In each material, the storage of fault structures was observed to approach saturation by $\gamma_P \approx 0.15$. Videos of the evolution of deformed microstructures with increasing plastic strain are made available in the online supplementary material. The deformation processes and morphologies presented in Fig. 2 are supported by tensile experiments of single crystal nanowires [38,48]. Although experimental studies inherently probe loading orientation factors (*i.e.*, Schmid effects), which are intentionally neglected herein, single-crystal nanowires represent a reasonable system for comparison to kMC simulations as they possess analogous boundary conditions (*i.e.*, free surfaces). Within this context, Ramachandramoorthy et al. [48] observed the nucleation and storage of thin twin lamellae ($< 5d_{111}$) in nanotensile experiments of single crystal Ag nanowires. Lee et al. [38] observed partial dislocation-mediated twinning and detwinning mechanisms in Au nanowires subjected to cyclic tensile/compressive loadings. These experimental reports correlate well with the deformation structures observed in kMC simulations.

The evolution of the fault fraction with increasing plastic strain for Ag, Au, and Cu is plotted in Fig. 3. The raw data from all replications of the kMC simulations is provided as normalized contours, as described above, and the averaged data is plotted using the relevant markers. Analytical predictions from the model defined in Section 2.2 are plotted in dashed stroke. In each material, a monotonic increase in the fault fraction is predicted, with Ag exhibiting the highest fault storage. The analytical model is in excellent agreement and captures the critical features of the kMC simulation data. In each material, the rate of increase in the fault fraction is highest at the incipient stages of plasticity and approaches a linear relation at higher plastic strains. Cu and Au exhibit only modest increases in the fault fraction after the early stages of plastic strain 0.01-0.05 plastic strain, whereas Ag tends to continue to store faults at higher strains, albeit at a reduced rate. This behavior is in line with the experimental literature, which reports a steep increase in the twin fraction during incipient plastic events [8]. In the case of Cu, a noticeable asymmetry exists in the binned kMC data, where the average response deviates significantly from the most common fault behavior. Since the fault fraction cannot be negative, this asymmetry arises due to the preference for the removal of fault structures in Cu over extended deformation, which produces frequent system configurations at $F = 0$.

Results for the evolution of the deformation twinning-accommodated strain are provided in Fig. 4. The agreement between model predictions and the kMC datasets is again excellent. As anticipated from the results of Fig. 3, Ag accommodates the most plastic strain by deformation twinning. Indeed, the evolution of $\gamma_F$ maps closely to the development of $F$, which reflects their correlation through the trailing nucleation probability term $P^-$ and the relations of Eqs. (4) and (6). Fig. 5 presents the evolution of the fault number density for each material in the study. Analytical predictions from the model defined in Section 2.2 are plotted in dashed stroke. For each material, the model predictions of fault number density match the kMC averages very closely. In each system, a monotonic increase in the fault number density is predicted as deformation proceeds. This behavior correlates well with the results of fault fraction and plastic partitioning calculations. Additional results showing the kMC data and model predictions for various fault density morphologies (*i.e.*, $n_k$) are provided in the online supplementary material.

The ability of the analytical model to capture the evolution of defect structures over extended plasticity is perhaps the most notable outcome of these results. Indeed, using only process barriers derived from the GPFE landscape, we have developed a methodology that can predict the partitioning of plastic strains between the mechanisms of deformation twinning and dislocation slip for several FCC materials. This physical description of strain partitioning is free from empirical fitting and is therefore anticipated to improve the phenomenological relations currently implemented in work hardening and crystal plasticity models. Examples, where a direct application of this approach would be beneficial, are found in the dislocation storage framework of Bouaziz and coworkers [25–27] and deformation-twinning crystal plasticity models [57–59], among others.

**3.2. Evolution of mechanism competition under extended deformation**

The excellent agreement between kMC simulations and analytical modeling motivates an examination of the evolution of mechanism competition in FCC metals over extended plastic deformation. Fig. 6 presents the evolution of the competition parameter, $\eta$, with the fault fraction. The analytical definition of $\eta$ (Eq. (10)) is plotted in dashed stroke for each material. The average kMC data is provided as markers and data distribution is represented as contours. As shown in the figure each of the materials exhibits a reduction in the competition parameter with increasing fault fraction. This finding is intuitive, as additional activation sites become available for trailing dislocation nucleation when more fault structures are present. The competition parameter is observed to be the largest for Ag across all fault fractions, whereas it quickly decreases towards parity for Cu and Au. The most common occurrences in the kMC data are well-captured by the analytical model for each material under study. Deviations of model predictions from the average kMC values are due to some infrequent, yet strongly favored leading nucleation configurations, which lead to asymmetries in the distribution of kMC data. These favored configurations arise due to rare circumstances where the Volterra fields of stored dislocations drastically reduce the barrier to nucleation of leading dislocations. Conversely, this effect is not seen in trailing dislocations, likely due to the co-planar positions of nucleation sites at faults formed by the action of leading dislocations.

Taking the nucleation of leading/trailing dislocations as representative of competition between deformation twinning and dislocation slip tendencies, the analytical model provides an excellent platform to compare against existing twinnability parameters in the literature. This comparison is particularly strong in incipient stages (*i.e.,* after nucleation of the first fault), where the system configurations across all parameters are similar. For this purpose, we have plotted $\eta$ against the twinnability parameters of Jo et al. [7], Jin et al. [6], Asaro and Suresh [5], and Tadmor and Bernstein [4] (Fig. 7). The incipient values of $\eta$ are plotted with markers, and the demarcation between twinning-dominated and slip-dominated behavior is denoted in a dashed stroke. Incipient data for Al has also been included to show alignment with other twinnability parameters in the slip-dominated regime. The incipient values of $\eta$ are calculated after the nucleation of an initial ISF (*i.e.,* $F = \frac{1}{N}, n_1 = n_F = \frac{1}{N}$, other $n_k = 0$). Comparison with the literature parameters yields several interesting results. For instance, the ranking of Ag as the material with the highest twinnability is common across all literature parameters. However, the sequencing of Au above Cu is unique to our calculations. This outcome finds support from several sources in the literature. For example, wire drawing experiments from English and Chin show that deformation twinning initiates at much lower stresses in Au than in Cu [60]. Indeed, this discrepancy in twinnability predictions and twinning stress data is noted in the seminal work from Tadmor and Bernstein [4]. We also note that $\eta$ returns the same conditional inequalities as other twinnabilty parameters when the microstructural evolution parameters are omitted and process barrier definitions are aligned. For instance, Jo et al. [7] defined twinnability process barriers using the relation $\frac{E}{\cos\theta}$, where $\theta$ is the angle between the Burger's vectors of dislocations. For $\theta = 60°$ (*i.e.,* conjugate leading/trailing partial dislocations) considered at comparable incipient conditions, $F = \frac{1}{N}, n_1 = n_F = \frac{1}{N}, 1 - F - 2n_F = 0$ with one leading nucleation site, and using the transformation $\gamma_{usf}^2 \approx \gamma_{usf}^1 + \frac{1}{2}\gamma_{isf}$ [6], the competition parameter reduces to $\frac{\gamma_{isf}}{\gamma_{usf}^1 - \gamma_{isf}} < 2$, which is the same inequality presented by Jo et al [7]. Through the development of this parameter, we have demonstrated a method to predict the twinnability of FCC metals by separately weighing the contributions of process barriers and deformation history towards the competition between deformation mechanisms. In a broad sense, this outcome expands the application of the twinnability concept to describe the evolution of deformation twinning and dislocation slip in deformed microstructures.

## 4. CONCLUSIONS

The competition between deformation twinning and dislocation slip has been studied for four common FCC metals (Ag, Au, Cu, and Al) using kMC simulations. In contrast to previous efforts, which examine only incipient events, the evolution of mechanism competition has been considered over extended plastic deformation. Kinetics in kMC simulations are informed directly by the critical features of the GPFE landscape and therefore provide an intrinsic comparison of mechanism competition. From the kMC simulation data, the evolution of the fault number density, fault fraction, and the partitioning of plastic strains between deformation twinning and dislocation slip mechanisms was measured. Results from these efforts show that Ag exhibited the highest storage of faults and the highest fault fraction over the entire deformation range studied. Based on kMC results, an analytical framework has been developed to provide a physical model for the mesoscale evolution of defect structures in FCC crystals. Predictions from this model find excellent agreement with kMC simulations. In addition, the relations of this model were used to define a competition parameter that can be used to examine the evolution of mechanism competition in FCC metals over extended deformation. Predictions from this parameter agree with experimental data showing the higher twinnability of Au relative to Cu, which is not captured by existing twinnability parameters. The outcomes of this study expand the applicability of deformation twinning theory beyond incipient plasticity and provide the community with relations for the evolution of fault fraction, fault number density, and strain partitioning between deformation twinning and dislocation slip mechanisms. These relations are free from empirical fitting constants and may be implemented to improve current work hardening and crystal plasticity models, which have previously relied on phenomenology.

## ACKNOWLEDGEMENTS

This material is based upon work supported by the National Science Foundation under Grant No. DMR-2144451. The authors gratefully acknowledge the Advanced Cyberinfrastructure for Education and Research (ACER) group at the University of Illinois at Chicago for providing computational resources and services needed to deliver research results delivered within this paper. URL: https://acer.uic.edu. The authors would like to thank G. Hibbard for useful comments and suggestions on the manuscript.

## AUTHOR CONTRIBUTIONS

R. J. performed the kMC simulations and contributed to the kMC code. M.D. conceived the project, developed the kMC code, performed some of the kMC simulations, and derived the analytical model. The manuscript was co-written by both authors.

## ONLINE SUPPLEMENTARY MATERIAL

Supplementary material is available in the online version of the paper or by email request from the corresponding author ([mattdaly@uic.edu](mattdaly@uic.edu)).

## CONFLICT OF INTEREST

On behalf of all authors, the corresponding author states that there is no conflict of interest.

**FIGURES**

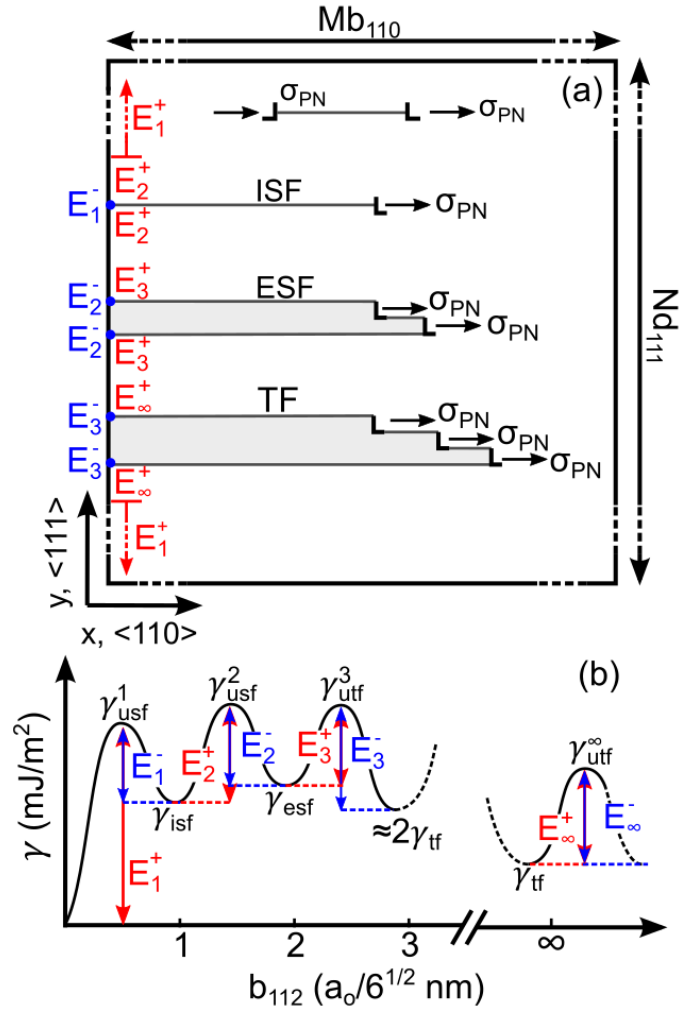

**Fig. 1:** (a) The kMC simulation cell with the relevant crystal directions and geometric parameters noted. The signature fault structures of deformation twinning (*i.e.,* ISF, ESF, and TF) and dislocation slip (i.e., a dissociated partial) are shown schematically. The nucleation barriers for deformation twinning and dislocation slip are shown in red and blue stroke, respectively. The determination of nucleation barriers is defined by the deformation history of the simulation cell. (b) The GPFE landscape for a typical FCC material. The relation between nucleation barriers and the local fault environment is illustrated. The values on the abscissa indicate the number of leading partial dislocations required to create each fault structure, where $b_{112}$ and $a_o$ are the magnitudes of the <112> Shockley partial dislocation Burger's vector and the lattice parameter, respectively. Please see the main text for a description of other symbols.

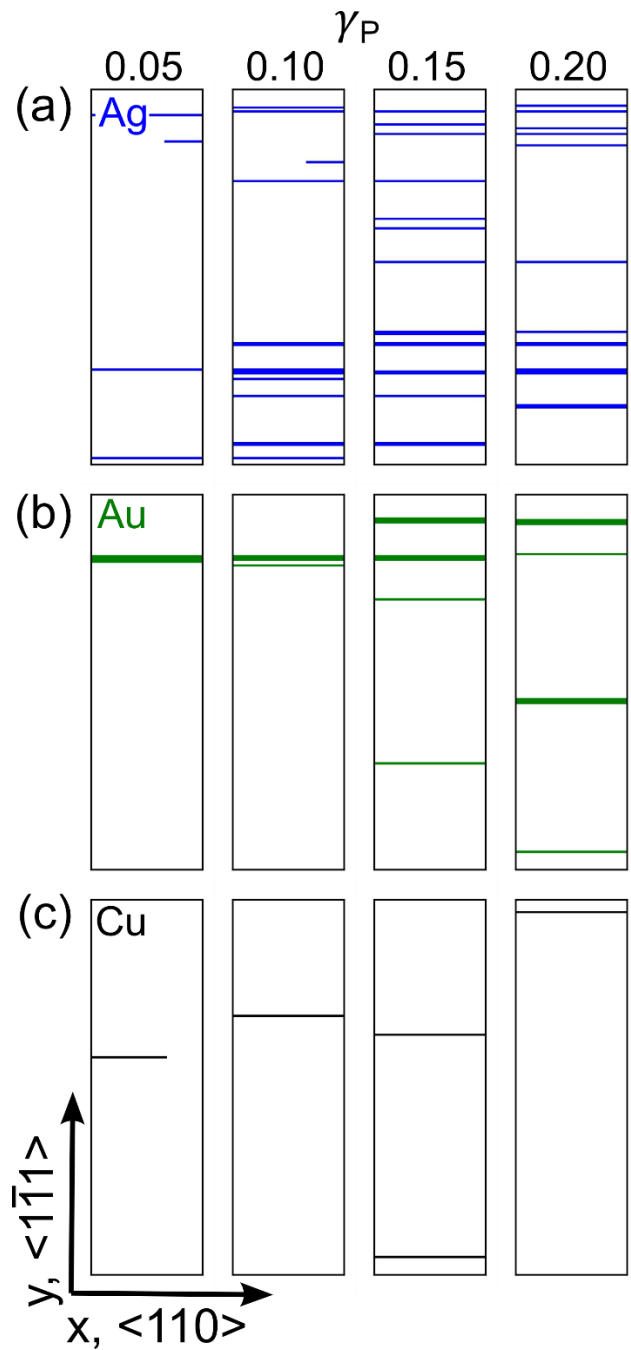

**Fig. 2:** Snapshots from kMC simulations at plastic strains of 0.05, 0.1, 0.15, and 0.2 for Ag (a), Au (b), and Cu (c). The shaded regions indicate the presence of a fault. For Ag, the simulation cell is segmented by several fault structures. By contrast, deformation in Cu is slip-dominated and planar faults are readily annihilated by the nucleation of trailing partial dislocations.

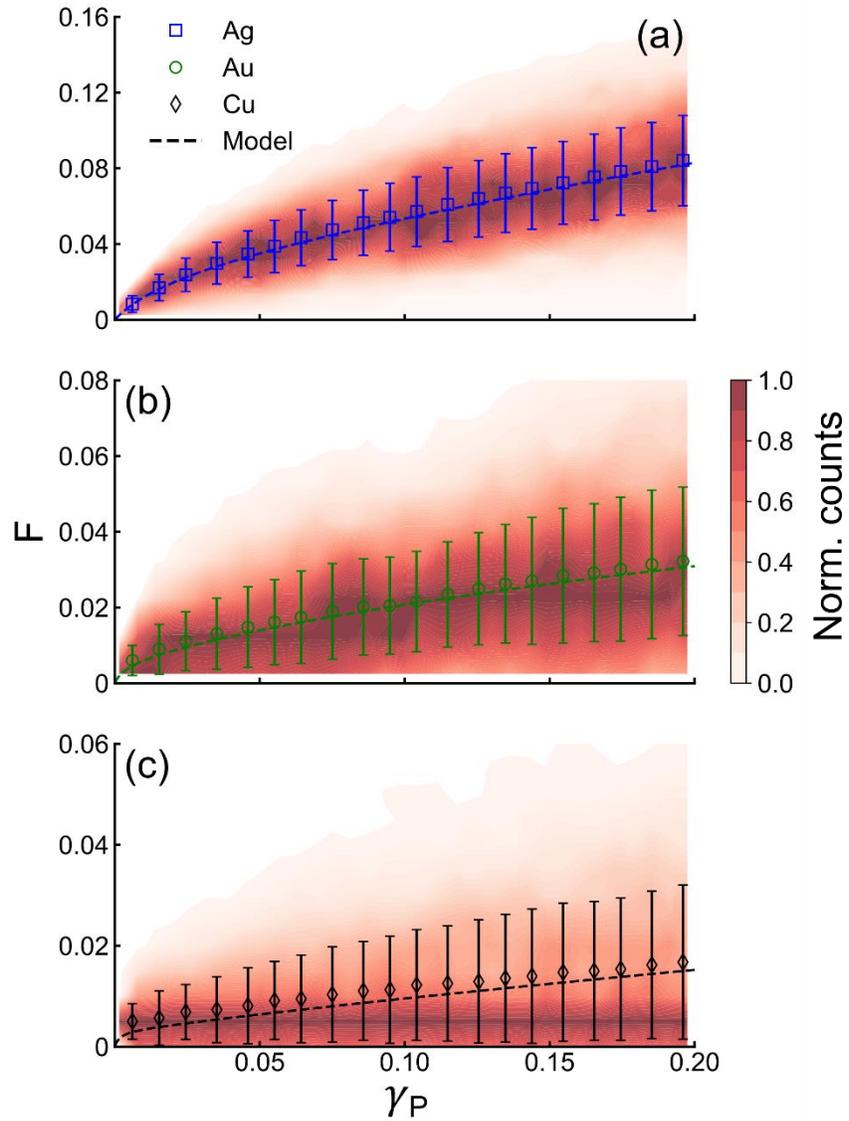

**Fig. 3:** The evolution of the fault fraction as predicted by kMC simulations is overlaid with the analytical model (dashed lines). The data is plotted for Ag (a), Au (b), and Cu (c). The average kMC data is plotted as markers. Error bars represent $\pm 1$ standard deviation over 500 replications of the kMC simulation. The contour plots are color-coded using a normalized counting scheme implemented along the ordinate axis. See the main text for further details.

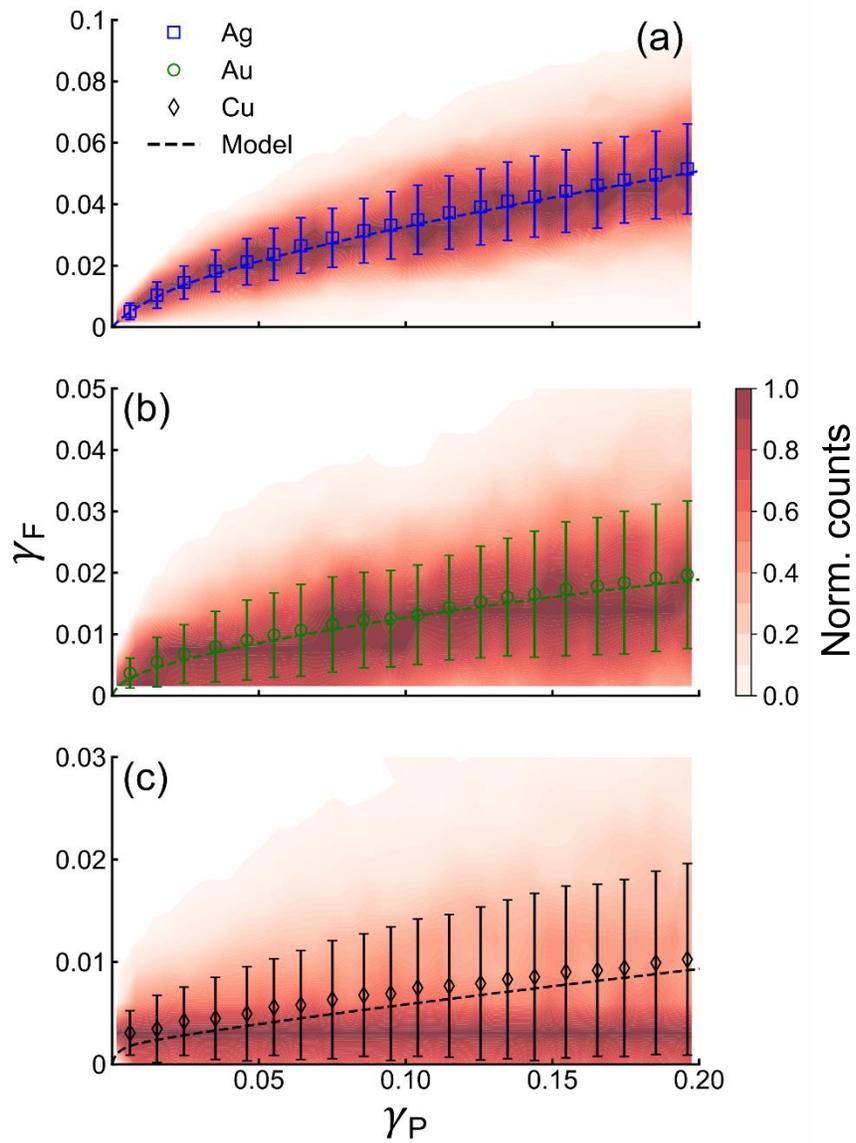

**Fig. 4:** The evolution of the plastic strain accommodated by twinning/fault formation over extended deformation for Ag (a), Au (b), and Cu (c). The average and standard deviation of kMC results are plotted in markers and error bars, respectively. The results of model calculations are overlaid in dashed stroke. The normalized counts of kMC replications are collected as described in the main text.

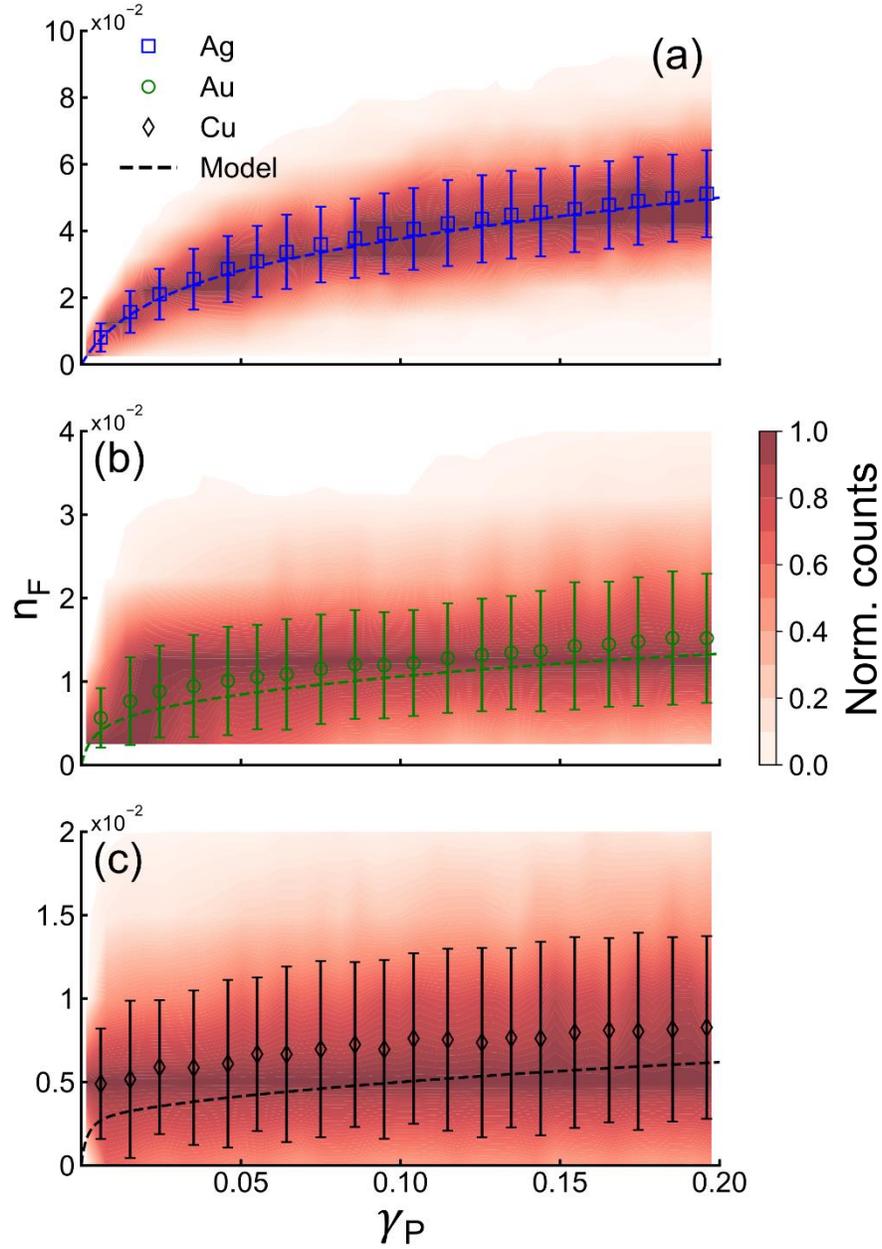

**Fig. 5:** The evolution of the fault number density as predicted by kMC simulations is overlaid with the analytical model (dashed line). The data is plotted for Ag (a), Au (b), and Cu (c). The average kMC data is plotted as markers. Error bars represent ±1 standard deviation over 500 replications of the kMC simulation. The raw kMC simulation data is shown in the contour plots. The contour plots are color-coded using a normalization scheme implemented along the ordinate axis. See the main text for further details.

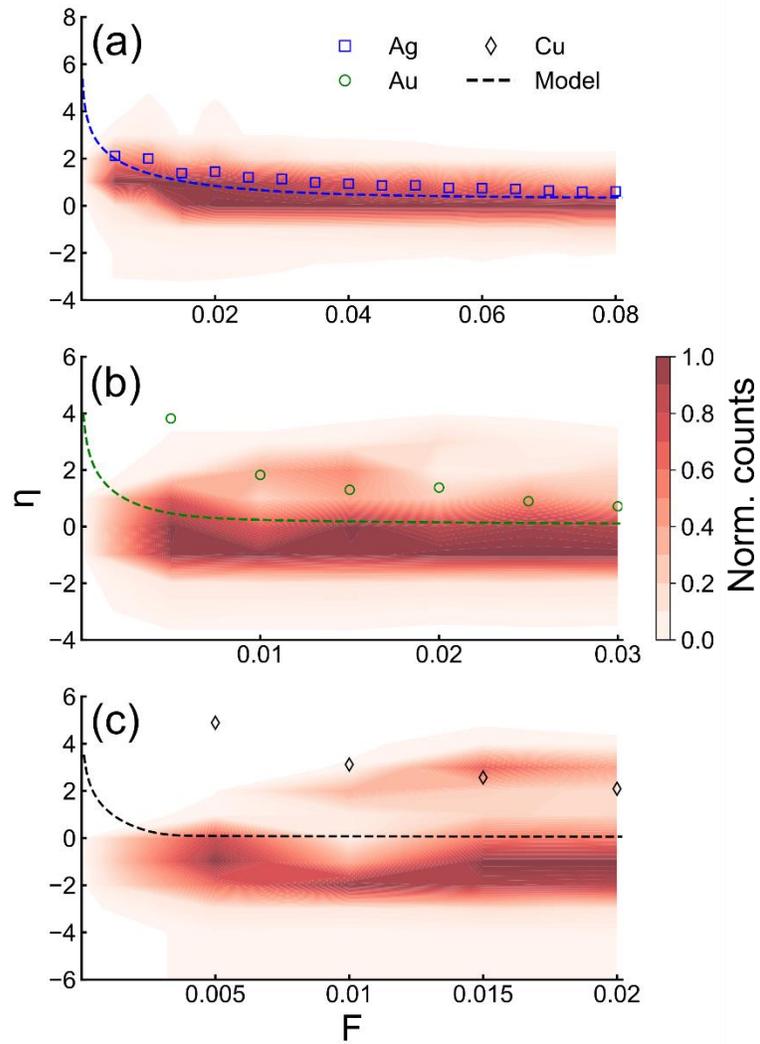

**Fig. 6:** Average kMC predictions (markers) for the competition parameter are overlaid with the analytical model (dashed line). The data is plotted for Ag (a), Au (b), and Cu (c). The contour plots show the raw kMC data and are normalized relative to the maximum bin values at increasing fault fraction intervals. See the main text for details. The results of the analytical model match well with the most common occurrence (contour data) in kMC simulations.

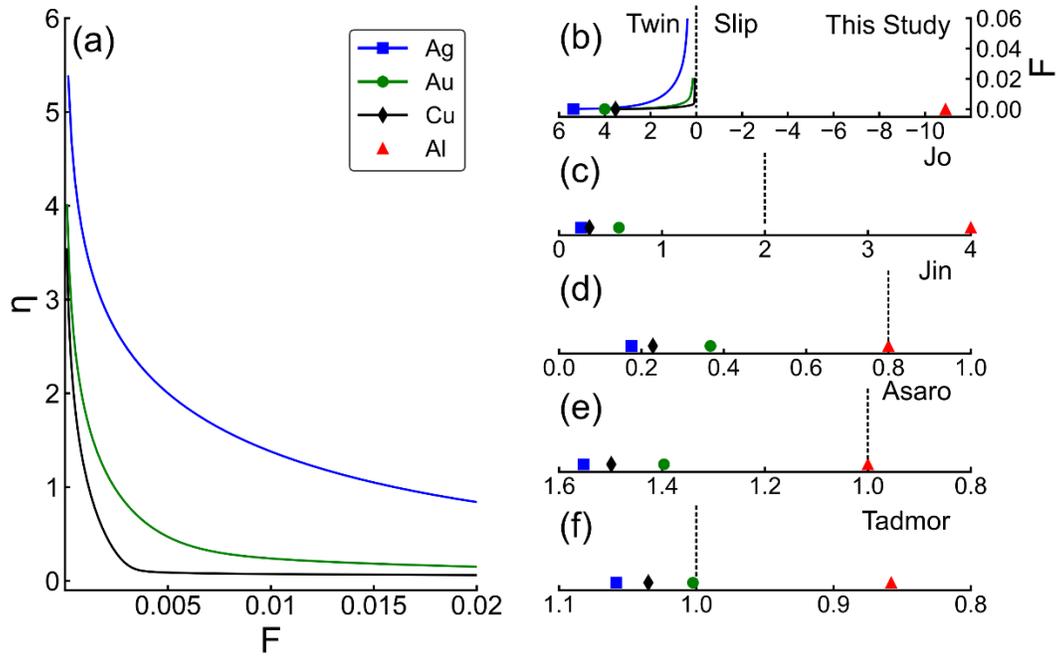

**Fig. 7:** (a) Model predictions for the evolution of the competition parameter over extended deformation is re-plotted for comparison purposes. (b) The competition parameter ($\eta$) is compared against the parameters of Jo et al. [7] (c), Jin et al. [6] (d), Asaro and Suresh [5] (e), and Tadmor and Bernstein [4] (f). The dashed line demarcates twinning- and slip-dominated regimes. Incipient data is plotted as a marker for each material and the incipient data for Al (slip-dominated) is provided for comparison.

# TABLES

**Table I:** Material parameters used in kMC simulations. Fault energies are provided in units of (mJ/m$^2$).[b]

| Material | $a_o$ (nm) | $G$ (GPa)[a] | $v$[a] | $R_o$ ($10^{13}$/s) | $\gamma_{usf}^1$ | $\gamma_{usf}^2$ | $\gamma_{utf}^\infty$ | $\gamma_{isf}$ | $\gamma_{esf}$ | $\gamma_{tf}$ |
|---|---|---|---|---|---|---|---|---|---|---|
| Ag | 0.409 | 27.8 | 0.43 | 3.94 | 91  | 100 | 93  | 16  | 12  | 8  |
| Au | 0.408 | 25.9 | 0.48 | 4.92 | 68  | 79  | 72  | 25  | 27  | 12 |
| Cu | 0.361 | 44.2 | 0.41 | 7.98 | 158 | 179 | 161 | 36  | 40  | 18 |
| Al | 0.405 | 25.9 | 0.36 | 9.66 | 140 | 196 | 135 | 112 | 112 | 50 |

[a]Calculated from compliance constants in Huntington [61] using the method of Bacon and coworkers [54,55].
[b]Retrieved from Jin et al. [6]